\documentclass[aps,prl,preprint,superscriptaddress]{revtex4-2}
\usepackage[final]{pdfpages}
\usepackage{graphicx}% Include figure files
\usepackage{dcolumn}% Align table columns on decimal point
\usepackage{bm}% bold math
\usepackage{color}

\makeatletter
\AtBeginDocument{\let\LS@rot\@undefined}
\makeatother

\begin{document}

%\preprint{}

\title{Formation of Anisotropic Polarons in Antimony Selenide}

\author{Yijie Shi}
 \affiliation{%
Wuhan National Laboratory for Optoelectronics, Huazhong University of Science and Technology, 1037 Luoyu Road, Wuhan, Hubei 430074, China}
 \affiliation{%
Advanced Biomedical Imaging Facility, Huazhong University of Science and Technology, Wuhan 430074, China}

\author{Xi Wang}
 \affiliation{%
Wuhan National Laboratory for Optoelectronics, Huazhong University of Science and Technology, 1037 Luoyu Road, Wuhan, Hubei 430074, China}
 \affiliation{%
Advanced Biomedical Imaging Facility, Huazhong University of Science and Technology, Wuhan 430074, China}

\author{Zhong Wang}
 \affiliation{%
Wuhan National Laboratory for Optoelectronics, Huazhong University of Science and Technology, 1037 Luoyu Road, Wuhan, Hubei 430074, China}
 \affiliation{%
Advanced Biomedical Imaging Facility, Huazhong University of Science and Technology, Wuhan 430074, China}

\author{Zheng Zhang}
 \affiliation{%
Wuhan National Laboratory for Optoelectronics, Huazhong University of Science and Technology, 1037 Luoyu Road, Wuhan, Hubei 430074, China}
 \affiliation{%
Advanced Biomedical Imaging Facility, Huazhong University of Science and Technology, Wuhan 430074, China}

\author{Fuyong Hua}
 \affiliation{%
Wuhan National Laboratory for Optoelectronics, Huazhong University of Science and Technology, 1037 Luoyu Road, Wuhan, Hubei 430074, China}
 \affiliation{%
Advanced Biomedical Imaging Facility, Huazhong University of Science and Technology, Wuhan 430074, China}

\author{Chao Chen}
 \affiliation{%
Wuhan National Laboratory for Optoelectronics, Huazhong University of Science and Technology, 1037 Luoyu Road, Wuhan, Hubei 430074, China}

\author{Chunlong Hu}
 \affiliation{%
Wuhan National Laboratory for Optoelectronics, Huazhong University of Science and Technology, 1037 Luoyu Road, Wuhan, Hubei 430074, China}
 \affiliation{%
Advanced Biomedical Imaging Facility, Huazhong University of Science and Technology, Wuhan 430074, China}

\author{Jiang Tang}
 \affiliation{%
Wuhan National Laboratory for Optoelectronics, Huazhong University of Science and Technology, 1037 Luoyu Road, Wuhan, Hubei 430074, China}

\author{Wenxi Liang}
 \email{wxliang@hust.edu.cn} 
 \affiliation{%
Wuhan National Laboratory for Optoelectronics, Huazhong University of Science and Technology, 1037 Luoyu Road, Wuhan, Hubei 430074, China}
\affiliation{%
Advanced Biomedical Imaging Facility, Huazhong University of Science and Technology, Wuhan 430074, China}%

\date{\today}

\begin{abstract}
Antimony Selenide (Sb$_2$Se$_3$) is an attractive candidate of photovoltaics with not yet satisfying efficiency. Beside defects, polaron formation originated from lattice distortion was proposed to account for trapping free carriers, and the subsequent photoexcitation dynamics and optoelectronic properties, but such a mechanism is still lack of structural observations. Here we directly track the pathways of carrier and lattice evolutions after photoexcitation through optical and electron diffraction pump-probe methods, revealing the temporal correlations between dynamics of both degrees of freedom. The observed opposite separation changes of Se2-Sb2 and Sb2-Sb1 atom pairs  in a few picoseconds, and the intermediate state induced by local structural distortions lasting several tens of picoseconds, coinciding with the optical phonons population and coupling, and the trapping process of carriers, respectively, together with the analyses of modulation on diffuse scattering by the atomic displacement fields of polaron model, indicate the formation of anisotropic polarons with large size. Our findings provide carrier and structural information for helping the elucidation of polaron scenario in Sb$_2$Se$_3$, and probably in materials with anisotropic structure and soft lattice which are popular in developing novel optoelectronics.
\end{abstract}

%\keywords{}

\maketitle


Polarons lay significant impacts on the carrier transport of materials constituted with deformable lattice with polarity\cite{ZhengEES2019,WuSAdv2021,FuCRev2023}, due to their origin of propagating charges immobilized by the lattice distortion field\cite{FranchiniNRM2021}. Theoretical and experimental efforts have been made to elucidate polaron-related phenomena over past decades, only the recently advanced time-resolved structural probes brought us the direct information in terms of lattice evolution in polaron dynamics, for example, the dichotomy of lattice displacements during polaron formation\cite{LiQMat2016}, the radially expanding strain field on nanometre-scale\cite{GuzelturkNMat2021}, and large and small polarons forming on different timescales\cite{CotretPNAS2022}, identifying their key roles in photoinduced processes determining the electronic and thermal conductivities. For the intense subjects of light conversion, further insights of structural response of polaron are desired, as many materials involved in novel optoelectronics possess soft lattice, in which polarons play a more important role in the carrier mobility which governs the device performances\cite{FuCRev2023}.    

Antimony selenide (Sb$_2$Se$_3$) is an attractive absorber material for photovoltaics\cite{ZhouNPho2015,WangJEC2018,LiNCom2019,TangNEn2020} with advantages of, e.g., nontoxic ingredients, proper bandgap, and high stability, but unsatisfying power-conversion efficiency\cite{WangNEn2017,YangNCom2019,DuanAM2022,ChenSRRL2022,SinghSEn2023} much lower than the theoretical prediction\cite{FilipPRB2013} regardless of the material and device engineering. Time-resolved measurements of photoemission, absorption and terahertz spectroscopies of carrier dynamics in nano-crystallite, polycrystalline thin film, and single crystal of Sb$_2$Se$_3$ and Sb$_2$S$_3$, a compound of same family with identical crystal structure and larger bandgap, demonstrated the localization of excited carriers\cite{WangJPCL2019,YangNCom2019,GradPRM2021}, which was attributed to the factor hampering device efficiency. Debates are raised on whether defects\cite{GradPRM2021} or an intrinsic mechanism of lattice deformation\cite{YangNCom2019} trapping carriers are the origin of deficit of open-circuit voltage loss, hence the limited efficiency of antimony chalcogenides. Both theoretical and experimental studies showed the complex of defect states in Sb$_2$Se$_3$\cite{HuangAAMI2019,StoliaroffAAEM2020,HobsonAPL2020}. Small polarons were proposed to account for the carrier trapping without saturation in Sb$_2$S$_3$\cite{YangNCom2019} and the anisotropic photoluminescence in Sb$_2$Se$_3$\cite{TaoASci2022}, but were denied by the measured reversal of anisotropic photoconductivity, which did not exclude large polarons\cite{LiuJPCL2022}. So far the observations of excited state dynamics in antimony chalcogenides were focused on carrier responses, information of structural response is no doubt helpful to elucidate the puzzling coupled dynamics of electronic and lattice degrees of freedom. 

In this work, we investigate the carrier evolutions through transient absorption (TA) spectroscopy, and lattice responses through ultrafast electron diffraction (UED) in thin films of polycrystalline Sb$_2$Se$_3$, picturing the formation of anisotropic polarons after photoexcitation. The separation changes of atom pair encoded in UED results are correlated with the optical phonons population and carrier trapping process captured in TA measurements, constructing the scenario of polaron formation dynamics. The calculation of diffraction intensity change implementing a Gaussian atomic displacement field of polaron model, matches well with the observed diffuse signals evolving in a specific crystallographic direction, suggesting the anisotropy and large size of polarons.\\ 

\textbf{Results}

The thin films of polycrystalline Sb$_2$Se$_3$ were grown through thermal evaporation (see Methods). Sb$_2$Se$_3$ possesses the quasi-one-dimensional (quasi-1D) lattice structure with orthorhombic unit cell, composed by (Sb$_4$Se$_6$)$_n$ ribbons stacking along the $c$-axis direction (space group of $Pbnm$) via covalent bonds and holding each other in the $a$- and $b$-axis directions via weak van der Waals forces, as depicted in Fig. 1a. An indirect bandgap of 1.25 eV is extracted from the Tauc plot of steady-state absorption spectra (see Supplementary Fig. 1). The experimental layouts of TA and UED are schematically illustrated in Fig. 1b, showing the spectroscopically resolved carrier dynamics recorded in the former and the diffraction patterns encoding transient lattice structure captured in the latter (see Methods), through pump-probe based measurements with above-bandgap excitation of 1.55 eV (800 nm).\\

\begin{figure}
\includegraphics{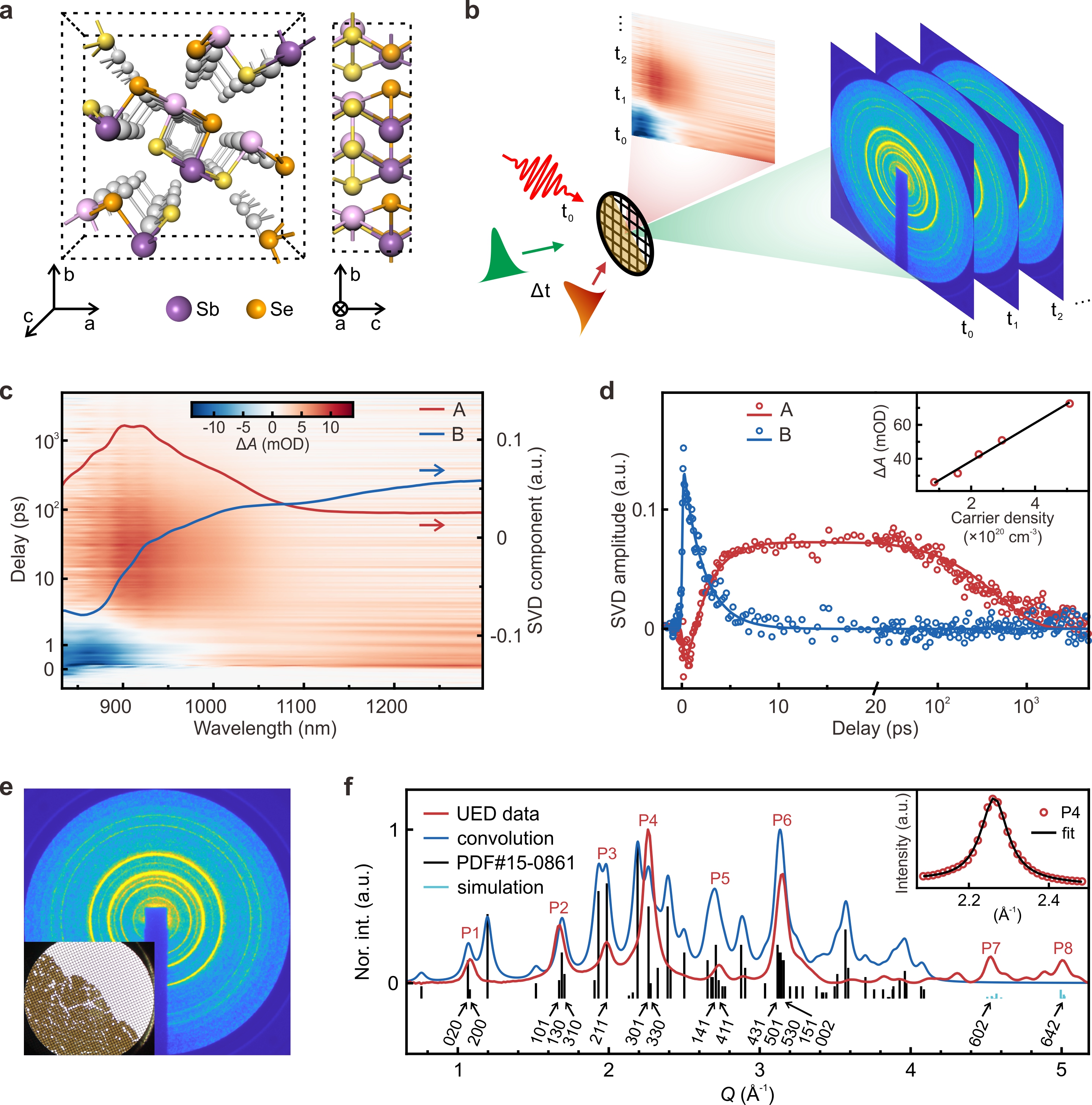}
\caption{\label{fig:figure1}Lattice structure and time resolved measurements. \textbf{a} Quasi-1D structure with orthorhombic unit cell (dashed frames). Note two layers in each Sb$_4$Se$_6$ unit due to the bonding orientations between Sb and Se atoms, denoted by different shades of colored balls. \textbf{b} Schematic layouts of TA and UED, measuring the transient states of carriers and lattice, respectively. \textbf{c} Pseudocolor contour plot of measured TA spectra, superimposed with two principle spectra extracted by SVD. OD, optical density. \textbf{d} Two principle kinetics extracted from \textbf{c} with global fitting. Inset: The maximum of trace A following the increase of injected carrier density. Black line, linear fit. \textbf{e} Electron diffraction pattern recorded by UED. Inset: Specimen of free-standing thin film. \textbf{f} Radially averaged distribution of diffraction intensity, showing Bragg peaks with possible indexes compared to calculated results (see Supplementary Note 2). $Q$, scattering vector. Inset: Lorentzian fit of the P4 peak profile.}
\end{figure}

\textbf{Localization of free carriers and population of coherent optical phonons.}
In TA measurements, we observe derivative-like featured spectral evolutions composited with a broad band photoinduced bleach (PB) signal in the wavelength range of 830--950 nm, evolving into a long lasting photoinduced absorption (PIA1) signal in a few picoseconds, and a broad band photoinduced absorption (PIA2) signal in the wavelength range of 950--1300 nm, decaying also in a few picoseconds, as shown in Fig. 1c the pseudocolor contour plot. The amplitude of PIA2 increases with a dependence on the wavelength to the power of $\sim$2.4 (see Supplementary Fig. 2), which is well ascribed as the intraband transitions of free carrier based on the Drude model\cite{YuBook2010}. Two principle spectra and associated kinetics are extracted by using the singular value decomposition (SVD), as depicted in Fig. 1c the superimposing spectra A and B, and in Fig. 1d the kinetic traces, corresponding to TA signals of PB/PIA1 and PIA2, respectively. The kinetics are analyzed using global fitting\cite{RuckebuschJPPC2012,StokkumBBA2004} combined with unbranched model\cite{NagleBioJ1982,RuckebuschJPPC2012} (see Supplementary Note 1 for details) yielding a rise time of 2.2 ps followed by a decay time of 475 ps for A, and a decay time of 2.2 ps for B. The temporal correlation of two species indicate the localization of free carriers, as reported in literatures\cite{WangJPCL2019,YangNCom2019,TaoASci2022}. We measured the carrier dynamics with increasing excitation fluences, confirming no saturation with injected carrier density up to 5.1$\times$10$^{20}$/cm$^2$, as depicted in the inset of Fig. 1d. Given the defects density of 10$^{14}$--10$^{17}$/cm$^2$ predicted by first principle calculations\cite{HuangAAMI2019,LiuPPho2017,StoliaroffAAEM2020}, and maximum 10$^{16}$/cm$^2$ measured in samples grown by various methods\cite{LuoAPL2014,ChenFOE2017,WenNCom2018,ChenSolR2021}, another origin besides defects is probably accounting for the localization of excited carriers in Sb$_2$Se$_3$.

Features of temporal oscillation during the early ten picoseconds are spotted by the zoomed-in inspection of TA results, as shown in Supplementary Fig. 3. We applied the fast Fourier transform and wavelet transform analyses on the oscillational signals, yielding three modes with frequency of 194 cm$^{-1}$ (M1), 119 cm$^{-1}$ (M2), and 42 cm$^{-1}$ (M3), respectively. These modes evolve with a cascade-like fashion, in which M1 and M2 populate then decay in less than 1 ps, while M3 rises when the former two are fading away then lasts in a period with lifetime of 2.6 picoseconds, see Supplementary Fig. 3d. Such oscillations were reported in TA measurements of single crystal Sb$_2$Se$_3$ and identified as phonon modes vibrating in $a$-$b$ plane and out-of-plane\cite{TaoASci2022}. The cascade-like temporal evolutions suggest the energy flow in the lattice subsystem, possibly mediating the atomic motions of lattice distortion, as discussed below. 

At this point we are able to construct the picture of carrier dynamics in Sb$_2$Se$_3$ after above-bandgap excitation. The excited carriers fill the edge of valence/conduction band with renormalized gap introduced by the indirect bandgap nature of Sb$_2$Se$_3$, with holes as the major species of carriers\cite{MavlonovSolE2020}, generating the broad band PB signal, while the free carriers absorb the photon energy of probe beam generating the PIA2. The band filling is soon followed by the localization of relaxed free carriers, originated from defects and probably another significant contributor, releasing plenty amount of available states for transitions from the valence/conduction band, hence generating the PIA1. At the early stage of evolutions, carriers deposit energy to the lattice through coupling to in-plane optical phonons, triggering the subsequent evolutions of atomic motions. 
\\
\\
\textbf{Anisotropic carrier-phonon coupling.}
In UED measurements, multiple Debye-Scherer rings (shown in Fig. 1e) are captured from the scattering of polycrystalline thin film attached to a copper grid as free-standing specimen (see the inset of Fig. 1e), allowing us for monitoring structural responses in various crystallographic directions. The patterns of each time delay are radially averaged using Hough transform to obtain Bragg peaks for corresponding lattice planes, as depicted in Fig. 1f. Although some of the lattice planes are not discerned in the measured profile due to the convolution of close peaks, limited coherence of the electron source, grain sizes, and possible texturing of the specimen, we are still able to extract the major features of structural evolutions by tracing the changes of Bragg peak. For simplicity, we label the obtained Bragg peaks as P1$-$P8, respectively. Each peak profile within a region of interest is fitted with Lorentzian function, as depicted in the inset of Fig. 1f, to extract quantitative information. 

All measured Bragg peaks show intensity drops upon excitation, with diffraction in some crystallographic directions changing apparently not scaling with the values of scattering vector, as depicted in Fig. 2a. The intensities of P1 (with a poor signal-to-noise rate probably due to the position of the ring next to the intense direct beam, see Fig. 1e) and P2 decrease more than those of P3 and P4, so does P5 compared to P6. We fit the kinetic traces within the early duration of several tens of picoseconds after excitation, finding that the transition times of intensity drop, which reflect the time spans for carriers coupling their excess energy absorbed from excitation photons to the lattice, vary distinctly in different crystallographic directions, as depicted in Fig. 2b. Both the asymptotes and the transition times of intensity drop address the anisotropic nature of Sb$_2$Se$_3$ lattice. 

\begin{figure}
\includegraphics{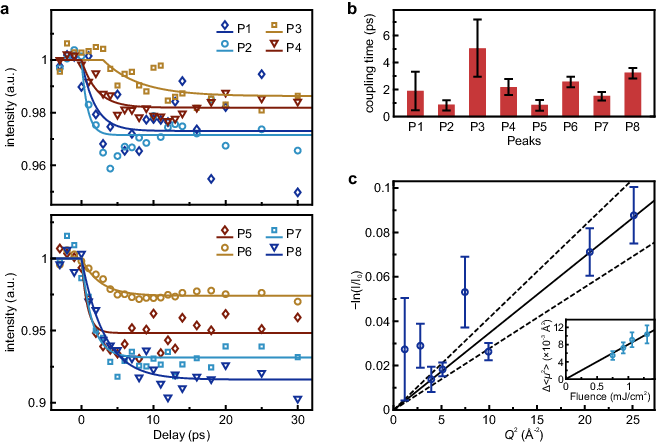}
\caption{\label{fig:figure2}Anisotropic structural dynamics. \textbf{a} Normalized intensity drops of P1-P4 (upper panel) and P5-P8 (lower panel) under excitation of 1.30 mJ/cm$^2$. Solid lines, monoexponential fits. \textbf{b} Transition times of intensity drop extracted from \textbf{a}, showing an anisotropic fashion of carrier-lattice coupling. \textbf{c} Intensity changes calculated from \textbf{a} as a function of the square of scattering vector at the delay time of 20 ps, showing that the evolutions of P1, P2, and P5 deviate from the DW model. Black line, linear fit for the DW effect. Dashed lines, enclosing the 95\% confidence interval for the linear fit. Inset, a linear dependence of the calculated changes of MSD, $\Delta \langle u^2\rangle$, on excitation fluences, indicating the thermalized lattice reaching a quasi-equilibrium state. Error bars in \textbf{b} and \textbf{c} represent the standard deviation of measurements.}
\end{figure}

One common contribution for diffraction intensity drop is the increase of mean square displacement (MSD) of atoms in the lattice, which is heated up by the energy transferred from excited carriers. Based on the well-known Debye-Waller (DW) model, the relationship of MSD and the change of diffraction intensity is described as
\begin{eqnarray}
I_{hkl}(t)/I_{hkl}(0)=e^{-\frac{1}{3}Q_{hkl}^2\Delta \langle u^2(t)\rangle},
\end{eqnarray} 
where $I_{hkl}(0)$ and $I_{hkl}(t)$ are the measured intensity of Bragg peak ($hkl$) before and after excitation, respectively; $\Delta\langle u^2(t)\rangle$ is the difference of MSD. We measured the structural dynamics with various excitation fluences to verify the thermalization of lattice. The calculated quantities of -ln($I(t)/I(0)$) (see Supplementary Note 3) for all Bragg peaks at 20 ps, as the lattice approaching a quasi-equilibrium sate, show that the intensities of the majority of peaks change linearly with the square of scattering vector, $Q^2$, in accordance with the harmonic assumption, while the changes of those of P1, P2, and P5 deviate profoundly from the linear trend, as depicted in Fig. 2c. Such deviations persist in results of all measured fluences, see Supplementary Fig. 4. We also calculated the quantities of $\Delta\langle u^2\rangle$, yielding a trend of linearity agreeing well with the DW model, as depicted in the inset of Fig. 2c. The intensity evolutions of P1, P2, and P5 suggest possible extra atomic displacements during the process of lattice thermalization, introducing nonthermal transient lattice distortions\cite{GuzelturkNCom2021,KrawczykJPCC2021}.
\\
\\
\textbf{Local structural distortions.}
The possible local distortions encoded in the observed intensity changes are better revealed by the analysis of pair distribution function (PDF)\cite{CockayneACSA1988,SiwickSci2003,MorrisonSci2014,ShiCPC2019} for the diffraction data. The calculated PDF $g(r,t)$ (see Supplementary Note 4 for details) with three peaks within a distance of 8 {\AA} is depicted in Fig. 3a, along with the distances of close atom pairs\cite{VoutsasZKCM1985} in the Sb$_4$Se$_6$ unit labeled in the inset. We assigned the peaks at 2.776 {\AA} and 4.179 {\AA} to the atom pairs of Se2-Sb2 and Sb2-Sb1, respectively, taking the peak positions and the distribution numbers of atom pair into account, see Supplementary Fig. 5 for details (the peak at 6.438 {\AA} is not discussed here because this value may correspond to multiple atom pairs which are indiscernible). 

\begin{figure}
\includegraphics{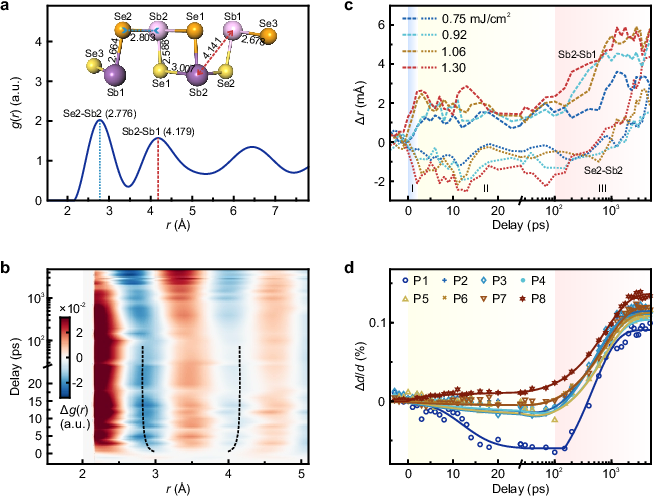}
\caption{\label{fig:figure3}Local lattice distortions and global lattice expansions. \textbf{a} Calculated PDF $g(r)$. The dotted line (cyan) and dashed line (red) indicate the peaks at 2.776 {\AA} and 4.179 \AA, corresponding to the distances of atom pairs of Se2-Sb2 and Sb2-Sb1 labeled in the inset, respectively. \textbf{b} Differential PDF $\Delta g(r,t)$ under excitation of 1.30 mJ/cm$^2$. Two dashed lines are the guides to the eye outlining the trend of changes of two peaks indicated in \textbf{a} during the early $\sim$100 ps. \textbf{c} Temporal changes of separations between the Se2-Sb2 pair (dotted lines) and the Sb2-Sb1 pair (dashed lines) under various excitation fluences. The separation changes evolve in stages marked with I, II, and III, respectively. \textbf{d} The relative changes of interplanar spacing $\Delta d/d$ under excitation of 1.30 mJ/cm$^2$, showing a two-stages process, see text.}
\end{figure}

We calculated the differential PDF by $\Delta g(r,t)=g(r,t)-g(r,t<0)$, which yields a distortion field reflecting the structural evolution in real space, as shown in Fig. 3b for the measurement under excitation of 1.30 mJ/cm$^2$. Both the negative signals (blue) in peak regions and the positive signals (red) in regions between peaks indicate the increase of MSD and the loss of order (see Supplementary Fig. 6 for schematic illustrations), consistent with the scenario revealed by the changes of diffraction intensity. The overall evolutions of the Se2-Sb2 and Sb2-Sb1 pairs show an opposite trend in a duration of several tens of picoseconds, as outlined by the dashed curves in Fig. 3b. We extracted the temporal separations, $\Delta r(t)$, of both atom pairs from the calculated $\Delta g(r,t)$ for all measured fluences, demonstrating an out-of-phase trend in the duration of early $\sim$100 ps and an in-phase trend in the duration from $\sim$100 ps to several nanoseconds, as depicted in Fig. 3c with the cubic spline interpolations\cite{WolbergCUCS1988} for smoothing the traces. The separation evolutions undergo three distinctive stages, which are denoted as I, II, and III, indicated by shaded regions with light colors of blue, yellow, and red, respectively, as shown in Fig. 3c. In stage I, the separation of Se2-Sb2 contracts with a length of $\sim$1.8 m{\AA} ($\sim$0.64$\%$) under excitation of 1.3 mJ/cm$^2$ in a few picosecond, resulting in a shift of $\Delta g(r,t)$ towards the $r-$ direction, while the separation of Sb2-Sb1 stretches with a length of $\sim$2.1 m{\AA} ($\sim$0.52$\%$), resulting in a shift of $\Delta g(r,t)$ towards the $r+$ direction. In stage II, the separations of both atom pairs stay constant (on the signal-to-noise level of our measurements), exhibiting a plateau state lasting almost 100 ps. In stage III, the separations of both atom pairs increase, as Sb2-Sb1 keeps stretching till it reaches the new equilibrium state after $\sim$1 ns, and Se2-Sb2 recovers to the steady state ($\Delta r=0$) in several hundred picoseconds then stretches to the new equilibrium state. The opposite ways of change followed by the synchronized-like movements of two atom pairs are indicative of a local lattice distortion preceding the global lattice deformation. 

These three-stage evolutions of atom pair separation are corroborated by the changes of lattice plane separation captured directly in the diffraction data. The peak positions extracted from the fitting of Bragg peak profiles, which are the reciprocals of the interplanar spacing, show temporal evolutions with two stages, as depicted in Fig. 3d with two shaded regions. During the stage of early $\sim$100 ps, all lattice planes except P1 (possibly affected by the direct beam) show negligible motions. In the stage of following time window, from $\sim$100 ps to several nanoseconds, all interplanar spacings increase. Given the temporal coincidence and time scales taken, we are confident that the synchronized-like stretches of atom pairs shown in Fig. 3c and the lattice expansions shown in Fig. 3d, in the durations of red shaded region, reflect the thermally driven global lattice deformations. The fits of kinetic trace for all lattice expansions yield time constants on the scale of several hundred picoseconds (see Supplementary Table 1), much larger than the several-picosecond durations for lattice heating up, demonstrating the soft lattice characteristics of Sb$_2$Se$_3$. In contrast to the long lasting thermal deformation, the preceding contraction and dilation of atom pairs rise and last in much shorter durations with no global lattice expansion accompanied, suggesting the nonthermal nature of the transient state of local distortions right after photoexcitation. 

We crosschecked the structural dynamics observed in Sb$_2$Se$_3$ by UED measurements of aluminum, a well studied material with trivial photoinduced dynamics of electron-phonon coupling and thermal expansion. The UED results with PDF calculations show straightforward evolutions of intensity drop due to lattice thermalization, uniform lattice expansions, and stretches of all atom pairs (see Supplementary Fig. 7 for details), as expected.   
\\
\\
\textbf{Modulations of polaron formation on diffuse scattering.}   
The strain fields originated from local distortions introduce changes of diffuse scattering intensity in the tail parts of Bragg peak, also known as Huang scattering\cite{HuangPRS1947,FultzBook2013}. We applied the analyses of Huang scattering for the measured Bragg peaks, found that the peaks with structural evolutions out of the DW description exhibit dynamics dependence on the scattering vector. The scattering vectors within a Bragg peak profile can be expressed as $\vec{Q}=\vec{G}+\vec{q}$, where $\vec{G}$ denotes the peak center, $\vec{q}$ denotes the scattering vector offset from $\vec{G}$, see the labeled blue curve in Fig. 4a. We took differentiation between the profiles collected after and before excitation, obtaining the normalized differential intensity $\Delta I/I_0$ with time- and $\vec{q}$- dependencies, as shown in Fig. 4a the pseudocolor contour plot for P5. The Huang scattering signal rise on both sides of the peak starting at the tail parts (regions with high $|\vec{q}|$ values). When the delay times increase, the starting points of rise move towards the peak center (regions with low $|\vec{q}|$ values), showing delayed responses with $\vec{q}$ dependence. Fig. 4b depicts the temporal traces of $\Delta I/I_0$ for each $|\vec{q}|$ with monoexponential fits, which yield quantified rise times depicted in Fig. 4c, showing a reciprocal dependence on $|\vec{q}|$. Bragg peaks with structural evolution following the DW description show no progressive rise of Huang scattering signal, see Supplementary Fig. 8 for a comparison of all measured peaks. The diffuse scattering signals introduced by lattice thermalization rise simultaneously over different scattering vectors, see Supplementary Fig. 9. Such thermally driven $|\vec{q}|$ independence is crosschecked by the UED measurements of aluminum, in which the diffuse scattering shows simultaneous rise, see Supplementary Fig. 10 for details.
  
\begin{figure}
\includegraphics{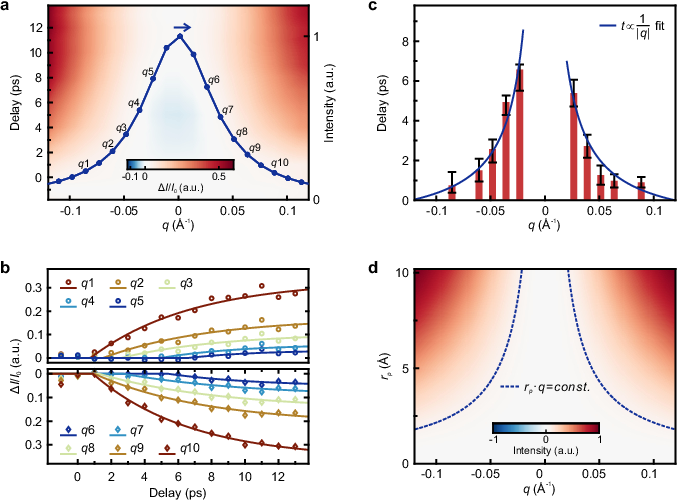}
\caption{\label{fig:figure5}Measured spatial and temporal evolutions of Huang scattering signals and calculation of polaron model. \textbf{a} Pseudocolor contour plot of the differential intensity of P5 over delay times and scattering vectors, with the normalized peak profile superimposed. Note that the depleting region (light blue) is inroduced by the intensity drop of Bragg diffraction, see Fig. 2a. \textbf{b} Temporal traces of $\Delta I/I_0$ with monoexponential fits for different $q$ denoted in \textbf{a}. \textbf{c} Rise times of Huang scattering signals for different $q$, extracted from \textbf{b}. Error bars represent the standard deviation of measurements. \textbf{d} Calculated intensity evolution of Huang scattering with the Gaussian polaron model, see text. Dashed lines outline the traces at where the products of $r_p\cdot{q}$ are constants.}
\end{figure}
  
Given the reciprocal relationship between scattering vector and dimension in real space, the progressive rises of Huang scattering signals map the temporal evolution of local  distortion to the spatial changes. Photoinduced local distortions have been observed in lead halides hybrid perovskites (MA)PbBr$_3$\cite{GuzelturkNMat2021} and thermoelectric material SnSe\cite{CotretPNAS2022}, and described by the formation of polaron which generates a Gaussian atomic displacement field given by
\begin{eqnarray}
\overrightarrow{u(r)}=A{\cdot}e^{-r^2/r^2_p(t)}\hat{r},
\end{eqnarray}
where $A$ denotes the amplitude of displacement, $r_p(t)$ denotes the time dependent radius of polaron, $\hat{r}$ denotes the direction of displacement vector. Subsequently, the normalized intensity with scattering vector dependence is given by
\begin{eqnarray}
\frac{\Delta{I(q,t)}}{I}\propto{A^2}(\vec{G}\cdot\hat{r})^2q^2r^2_p(t)\cdot{e}^{\frac{q^2r^2_p(t)}{2}},
\end{eqnarray}
where $\vec{G}$ and $q$ are above defined. We calculated the scattering intensities with contributions from the distortions described by this model (see Supplementary note 5 for details), yielding results which well reproduce the measured Huang scattering signals of P5, as shown in Fig. 4d, manifesting the formation of polarons as the origin of observed local distortions. Taking the reciprocal relationships between the rise times and $|\vec{q}|$, as well $r_p$ and $|\vec{q}|$ into account, we phenomenologically have $r_p\approx{kt}$, which yields $k=3.4$ \AA/ps by fitting the separation changes of atom pairs obtained in Fig. 3c with $r_p$ substituted into the polaron model of Equation (2), see Supplementary Fig. 11 for details. The progressing process of Huang scattering signal rise ceases at $\sim$10 ps, as shown in Fig. 4a and 4b, thus we have an estimated half high full width of $2\sqrt{2ln2}r_p\approx$ 8 nm for the lower bound of polaron size, which is much bigger than the dimension of Sb$_2$Se$_3$ unit cell ($a=11.633$ \AA, $b=11.78$ \AA, and $c=3.985$ \AA). The modulations on diffuse scattering are observed only in specific crystallographic directions in our measurements, indicating the anisotropy of polarons formed in Sb$_2$Se$_3$, in contrast to the reported spherically isotropic polarons in (MA)PbBr$_3$\cite{GuzelturkNMat2021}.
\\
\\
\textbf{Discussion}

Both defects, intrinsic or damage introduced\cite{EhrhartJNM1994}, and the formation of polarons through carrier trapping by defects or carrier self-trapping by excitation\cite{FranchiniNRM2021} generate local distortion fields in lattice. The direct effects by defect, which are irreversible for photoexcitation, can be excluded for pump and probe measurements. We can also exclude the significant contribution of polaronic effects originated from carrier trapping by defects, for the following assumptions against our observations: (i) Intrinsic defects are supposed to be saturated by photodoped carriers with density on a scale of $\sim$$10^{17}$ cm$^{-3}$; (ii) Sb$_2$Se$_3$ possesses rich type and amount of intrinsic defects resulting in various recombination centers at different atom sites distributed randomly\cite{HuangAAMI2019,StoliaroffAAEM2020,LiuPPho2017}, hence the generated signals of distortion field measured by UED with polycrystalline specimen are supposed to be isotropic, and more sensitive in regions with large scattering vectors. Furthermore, our observations reveal the local distortions around the site of Sb2 in (Sb$_4$Se$_6$)$_n$ ribbons, in agreement with the reported calculation results of distortions around the site of five-coordinated Sb2 binding the injected electrons, as well the contraction of Se2-Sb2 bond\cite{TaoASci2022}. The consideration of photoinduced lattice distortions accounting for trapping free carriers in  Sb$_2$Se$_3$ thus stands. One more feature should be taken into account is the soft lattice characteristics of Sb$_2$Se$_3$ demonstrated by the observed long duration of lattice thermal expansions, consistent with the predictions of small isotropic shear modulus in Sb$_2$Se$_3$\cite{KocSSS2012}, and carrier self-trapping preferring in materials with soft lattice\cite{MartinPRB1997}.

The coincided carrier trapping processes observed in TA and structural evolutions observed in UED construct the picture of photogenerated polaron formation. Although Sb$_2$Se$_3$ is nonpolar\cite{PetzeltFerr1973}, we speculate that the local distortions with asymmetric separation changes (on the signal-to-noise level of our measurements) of atom pairs rising within 2 ps, a duration for the A$_g$ phonons in the $a$-$b$ plane losing their coherence and coupling to the B$_g$ phonons in the out-of-plane direction through anharmonic processes\cite{TaoASci2022}, probably introduce an intermediate state with long-range polarity which interacts with the excited carriers and populated optical phonons, forming polarons through Fröhlich coupling\cite{FranchiniNRM2021}. The projection of modulation on diffuse scattering profound only in certain crystallographic directions combining the estimated large size, are consistent with the theoretical calculations which predicted anisotropic large polarons in quasi 1-D solids, resulting in charge movements as thermally activated hopping in the directions of inter-ribbon but as quasi-free particles along the ribbons\cite{HolsteinMCLC1981}, and the experimental measurements which reported the anisotropic transport property of temperature dependent photoconductivity\cite{LiuJPCL2022}. We note that the values of separation change of atom pairs in our results are much smaller than those predicted by calculations\cite{TaoASci2022}. Three points are noteworthy: (i) The separation changes obtained by differential PDF calculations are the sum result of lattice disorder and bond change and, as illustrated in Supplementary Fig.6. (ii) The strain fields introduced by polarons are spatially inhomogeneous, but the scattered intensities contributed to the signals of a specific Bragg peak are actually spatially averaged, hence the derived PDF signals. (iii) The intensity of a Bragg peak is contributed by all atoms in the probed volume, but the signals of separation changes of atom pair are contributed by atoms only involved in the polarons. 

The phonon dressing of polarons alters both the mobility and effective mass of moving carriers, subsequently the electric and thermal transporting properties in materials. Our findings bring in the information of both carrier and lattice responses for better understanding the polaron formation in highly anisotropic materials, and hopefully for harnessing these quasiparticles in applications of transforming and transferring energy.
\\
\\
\textbf{Methods}
\\
\textbf{Specimen preparation}. The thin films of polycrystalline Sb$_2$Se$_3$ were grown on the substrate of freshly cleaved single crystal KCl through thermal evaporation in a vacuum chamber with pressure of 4$\times$10$^{-6}$ Torr, then annealed at 200 {\textcelsius} in a glove box for 40 minutes. The thickness of films were grown to 22 nm calibrated by an in situ quartz crystal thickness monitor. Each batch of specimens were divided into two groups, one group with KCl substrate for measurements by TA, the other group loaded on TEM grids with substrate removed by deionized water for measurements in UED.   
\\
\textbf{Transient absorption spectroscopy}. The TA measurements were performed under the ambient atmosphere through an Helios spectrometer (Ultrafast System), pumped by 800 nm femtosecond pulses with pulse width of 35 fs and repetition rate of 2.5 kHz. The radius of pump and probe spots were 100 $\mu$m and 30 $\mu$m, respectively. The instrument response function of Helios was measured as 72 fs.
\\
\textbf{Ultrafast electron diffraction}. The UED measurements were performed under the ultrahigh vacuum condition through a homebuilt ultrafast electron diffractometer with estimated temporal resolution less than 1 picosecond\cite{HuCPB2021}, pumped by 800 nm femtosecond pulses with pulse width of 35 fs and repetition rate of 5 kHz. The probe electrons were accelerated by an electric field of 30 kV, generating diffraction patterns through a transmission geometric setup. Each pattern was recorded with an intensified EMCCD with 15000 pulses accumulated. The radius of pump and probe spots were $\sim$200 $\mu$m and $\sim$50 $\mu$m, respectively.  
\\
\\
\textbf{Acknowledgements}
\\
We thank the Analytical and Testing Center in Huazhong University of Science and Technology for the supports.
\\
\\
\textbf{Author contributions}
\\
W.L. conceived of and supervised the project. Y.S. performed the measurements with supports from X.W., Z.W., Z.Z., F.H., and C.H. C.C. prepared the samples under supervisions from J.T. Y.S. and W.L. analyzed the data with discussions with X.W., Z.W., Z.Z., F.H., and C.H. Y.S. and W.L. wrote the manuscript with contributions from all authors.
\\
\\
\textbf{Competing interests}
\\
The authors declare no competing interests.

\bibliography{Sb2Se3-polaron.bib}


\section*{Supplementary material}

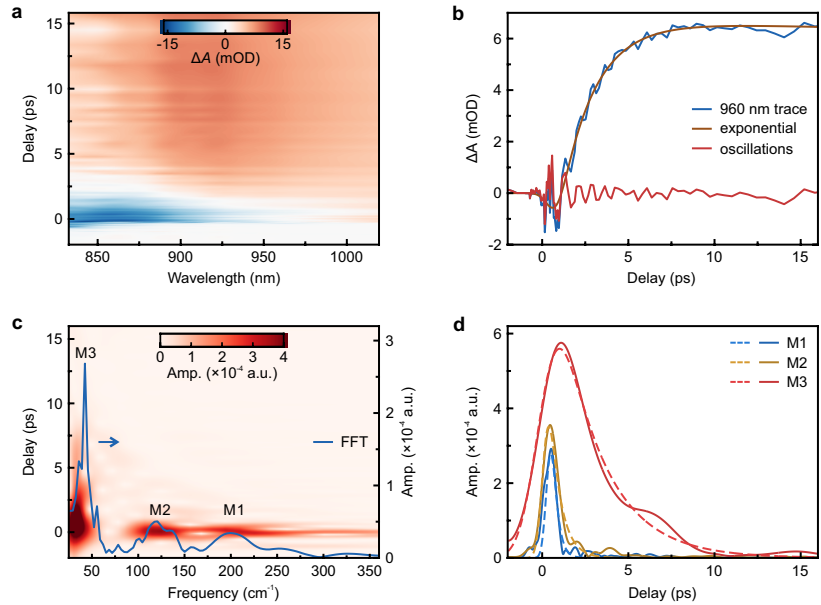

**Supplementary Figure 3.** Oscillation signals in the early stage of TA results. **a** Pseudocolor contour plot of TA spectra within 15 ps after excitation. **b** Prominent oscillations superpositioning on the kinetic traces (blue) at 960 nm. The oscillations (red) are clear with the exponential component (brown) subtracted using a moving average filter. **c** Frequency spectrum (blue trace) obtained by Fourier transform of the oscillation trace in **b**, superimposed on the wavelet transform results (pseudocolor contour plot). **d** Temporal traces of M1, M2 and M3 (solid lines) extracted from **c** and the exponential fits (dashed lines). The decay times of M1 and M2 are 283 fs and 387 fs, respectively. M3 rises in 925 fs then decays in 2.6 ps.

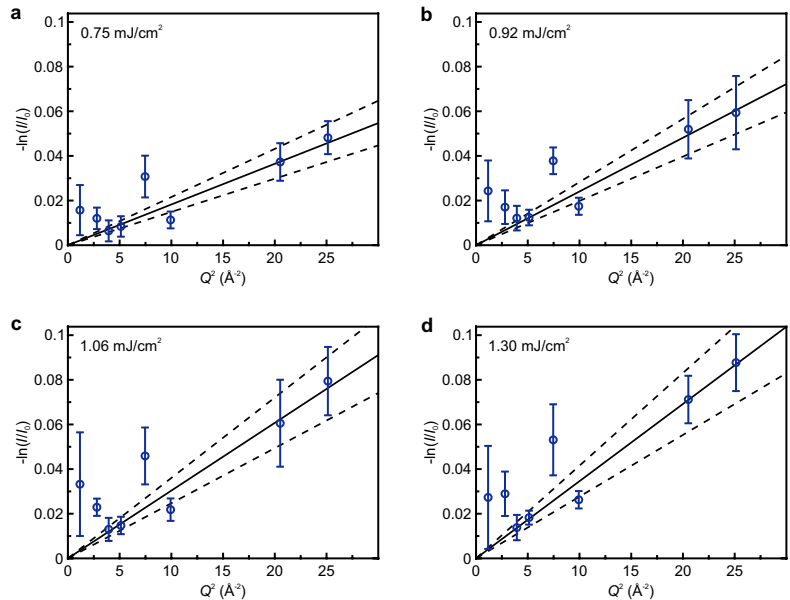

**Supplementary Figure 4.** Quantities of -ln($I/I_0$) at delay time of 20 ps as a function

of $Q^2$ excited with different fluences, showing the intensity changes of P1, P2 and P5 out of the DW model under harmonic assumption. Black lines, linear fits with zero intercept for the DW effect. Dashed lines, enclosing the 95% confidence intervals of the linear fit.

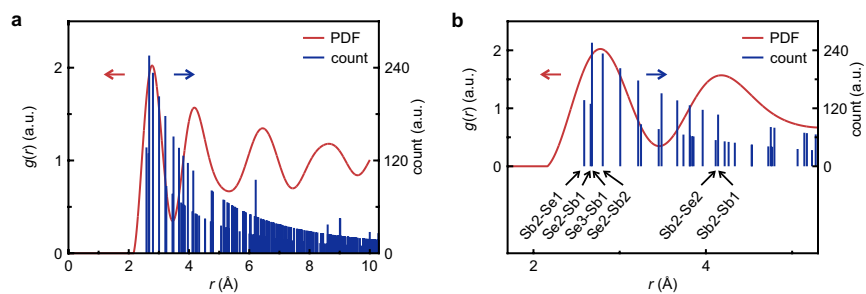

**Supplementary Figure 5.** Assignment of PDF peaks. **a** Calculated PDF and the distribution numbers of atomic pair counted within a sphere with a radius of 12 Å. **b** Zoomed-in inspection for the range of 5 Å in **a**. Taking the peak positions and the distribution numbers into account, the peaks located at 2.776 Å and 4.179 Å are assigned to atom pairs of Se2-Sb2 and Sb2-Sb1, respectively, labeled in the inset of Fig. 3a in the main text.

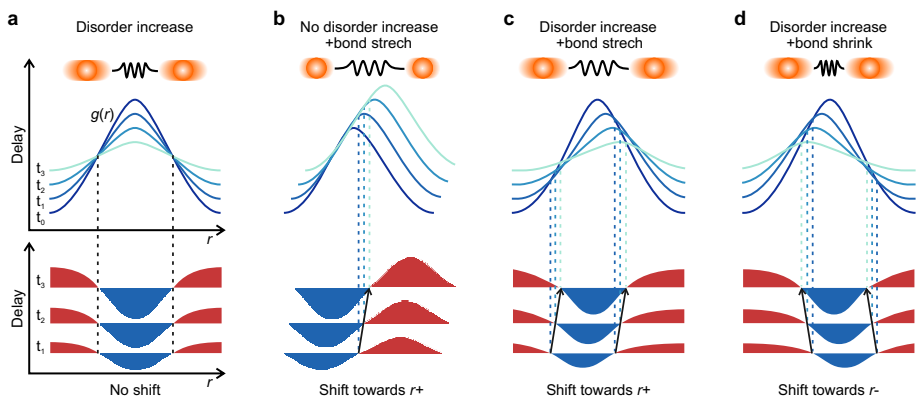

**Supplementary Figure 6.** Schematic diagrams for four possible situations of differential PDF signal, combining the changes of disorder and bond length. Note that the effect of disorder change competes with that of bond length change.

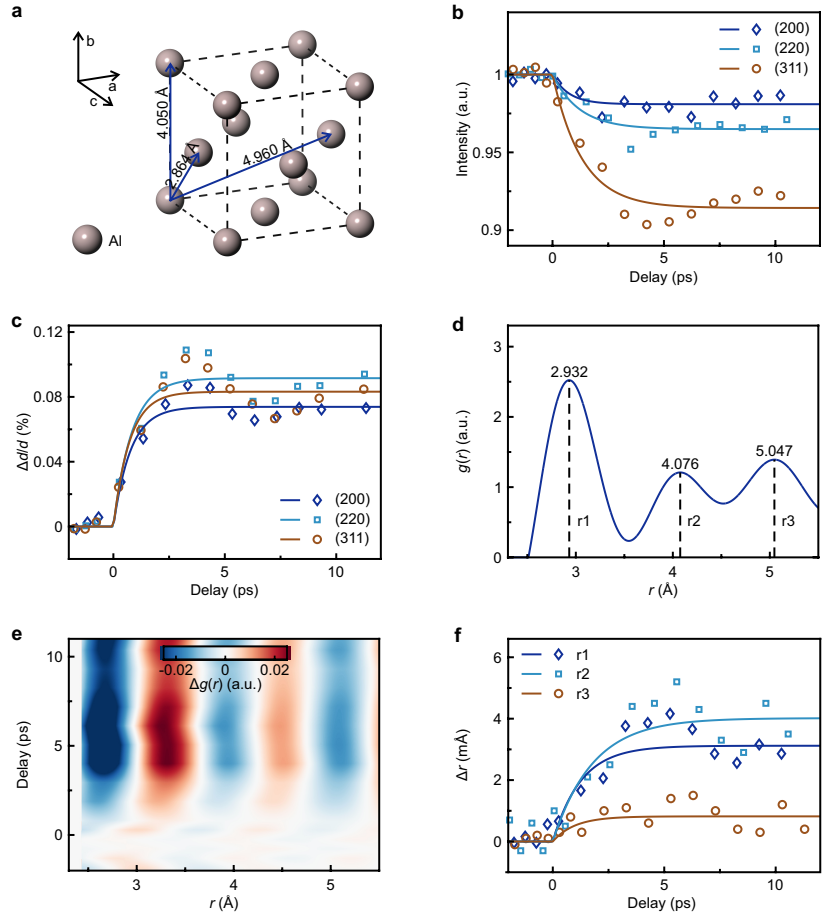

**Supplementary Figure 7.** Structural dynamics of aluminum thin films. **a** Face-centered cubic crystal lattice with three atom pairs of interests labeled. The lattice constants for a unit cell are $a=b=c=4.05$ Å. **b** Intensity drops, and **c** expansions of interplanar spacing measured in Bragg peaks of (200), (220), and (311). Note that the oscillatory responses superimposed on the exponential evolutions are introduced by the motions of breathing mode. **d** Calculated PDF with three peaks corresponding to atom pairs labeled in **a**. **e** Differential PDF, showing all peaks shifting towards the $r+$ direction when the disorder increases accompanied with atom pair distances increase. **f**, Temporal traces of $\Delta r$ extracted from **e**, showing only expansions as results of lattice thermalization.

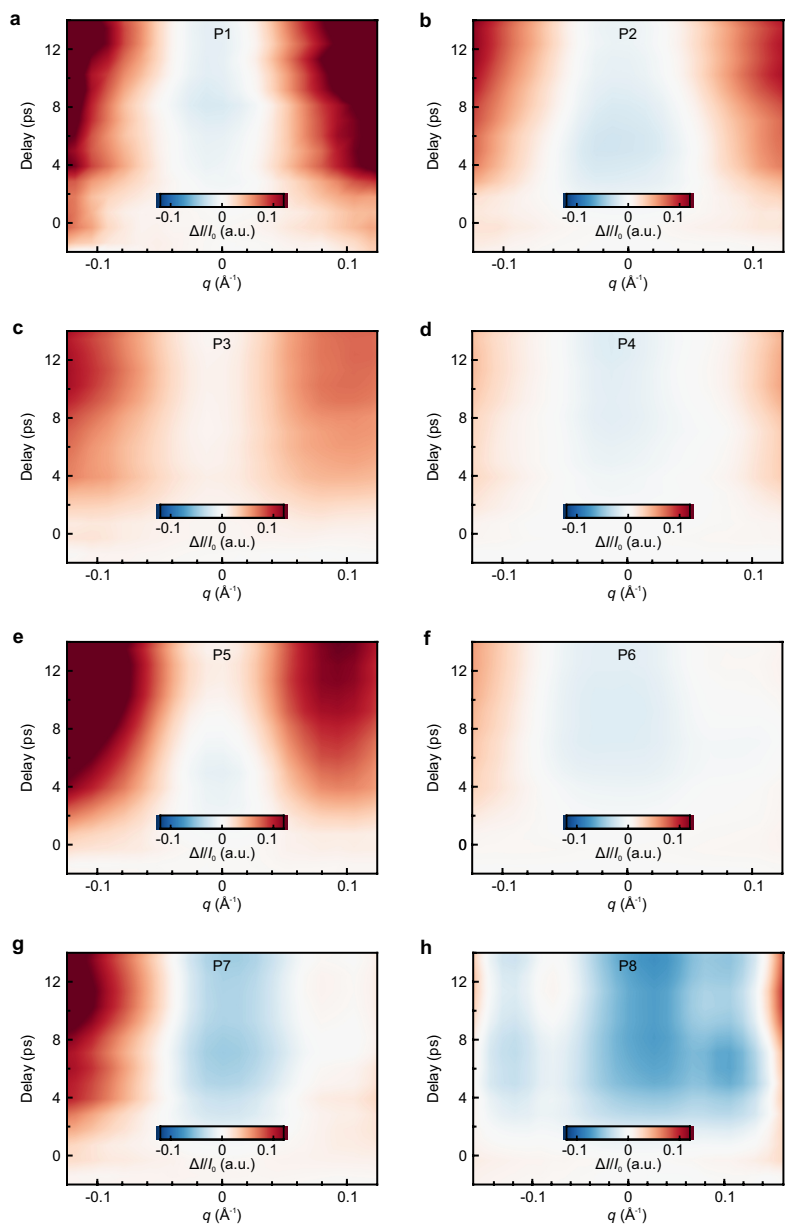

**Supplementary Figure 8.** Comparison of diffuse scattering signals for all measured Bragg peaks of $Sb_2Se_3$. P1 and P2 show progressive rises like P5, but with low signal-to-noise ratio. P3, P4, P6, P7, and P8 show simultaneous rises as results of lattice thermalization.

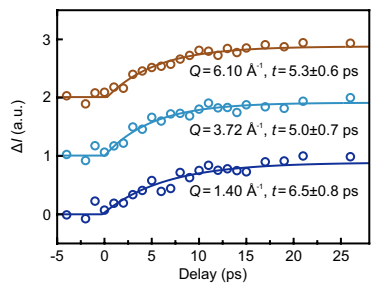

**Supplementary Figure 9.** Simultaneous rises of diffuse scattering signals over different scattering vectors $Q$=1.40, 3.72, and 6.10 Å$^{-1}$, which are far away from Bragg peaks. $t$, fitted rise times.

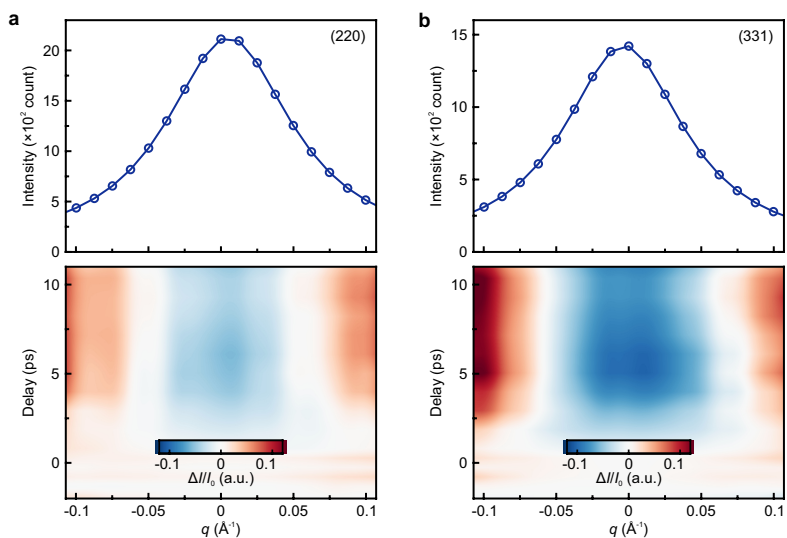

**Supplementary Figure 10.** Differential diffraction intensities of peak (220) (**a**), and (311) (**b**) of aluminum, showing simultaneous rises of diffuse scattering signals as results of lattice thermalization. Upper panels, peak profiles. Lower panels, pseudocolor contour plots.

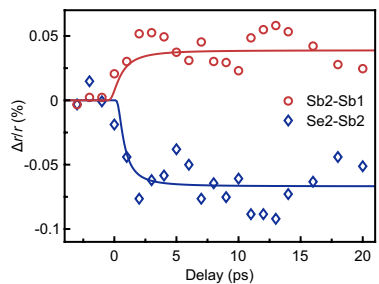

**Supplementary Figure 11.** Fitting separation changes of Se2-Sb2 and Sb2-Sb1 atom pairs, with substituting $r_p = kt$ into the polaron model $u = A \cdot \exp(-r^2/r_p^2)$. Taking

the Sb2 atom as the center of the forming polaron, $u$ equals $\Delta r$ at this point, hence we are able to fit the measured $\Delta r/r$ to estimate the lower bound of $r_p$. The fitting results of both traces yield $k \approx 3.4$ Å/ps.

**Supplementary Table 1.**

Table 1. Fitted expansion times for traces in Fig. 3d in the main text.

| Peak | P1 | P2 | P3 | P4 | P5 | P6 | P7 | P8 |
|---|---|---|---|---|---|---|---|---|
| Expansion time (ps) | 464 | 580 | 510 | 693 | 760 | 748 | 786 | 780 |

**Supplementary Note 1: Global fitting**

The rise of A component accompanied with the decay of B component, extracted by SVD, conforming to the unbranched model[1, 2]. We applied triple concentration components, which are labeled as $S_1$, $S_2$, and $S_3$, representing the convolution of the exponential attenuation term and the instrument response function described in a triple-compartment model, to implement the global analysis for the two kinetics. $S_1$, $S_2$, and $S_3$ are defined as:

$$\begin{cases} S_1(t) = \exp(-k_1 t) \oplus \text{IRF}(t) \\ S_2(t) = \dfrac{k_1}{k_2 - k_1}[\exp(-k_1 t) - \exp(-k_2 t)] \oplus \text{IRF}(t) \\ S_3(t) = k_1 k_2 \left[ \dfrac{\exp(-k_1 t)}{(k_2 - k_1)(k_3 - k_1)} + \dfrac{\exp(-k_2 t)}{(k_1 - k_2)(k_3 - k_2)} + \dfrac{\exp(-k_3 t)}{(k_1 - k_3)(k_2 - k_3)} \right] \oplus \text{IRF}(t) \end{cases}, \quad (1)$$

where $k_1$, $k_2$, and $k_3$ are the decay rates of each compartment, respectively. The instrument response function is generally described by a Gaussian function:

$$\text{IRF}(t) = \dfrac{1}{\sigma\sqrt{2\pi}} \exp\left[\dfrac{(t-\mu)^2}{2\sigma^2}\right], \quad (2)$$

where $\mu$ is the time zero, $\sigma$ represents the time width. The kinetics of A and B are the superpositions of concentration components with different proportional coefficients, so we have:

$$\begin{bmatrix} a(t) \\ b(t) \end{bmatrix} = \begin{bmatrix} C_{11} & C_{12} & C_{13} \\ C_{21} & C_{22} & C_{23} \end{bmatrix} \times \begin{bmatrix} S_1(t) \\ S_2(t) \\ S_3(t) \end{bmatrix} \quad (3)$$

where $C_{mn}$ are the proportional coefficients. The decay rates $k_1$, $k_2$, and $k_3$, which are

the reciprocal of the decay times, can be obtained by fitting the above kinetic matrix.

**Supplementary Note 2: Indexing Bragg peaks**

The standard XRD data of Sb$_2$Se$_3$ (black lines in Fig. 1f in the main text, PDF Card No.:15-0861, space group: Pbnm#62) provide peak information for the diffraction angle over 0 to 30°, which is equal to $0 \leq Q \leq 4.08$, $Q = 4\pi sin\theta/\lambda$. The blue curve depicts the convolution of a Lorentzian profile with the standard XRD data. The cyan lines in Fig. 1f in the main text depict the possible diffraction peaks, obtained through XRD simulation (VESTA) near the measured P7 and P8, which are out of the range of scattering vector provided by the PDF Card.

**Supplementary Note 3: Debye-Waller description**

The attenuation of diffraction intensity originated from atomic motions can be described by the Debye-Waller model[3]:

$$I_{hkl}(t) = I_{hkl}^0 \exp\left[-\frac{1}{3} Q_{hkl}^2 \langle u^2 \rangle(t)\right], \tag{4}$$

where $\langle u^2 \rangle(t)$ is the time dependent mean square displacement of atoms. Equation (4) gives the linear dependence between $-\ln(I_{hkl}^{norm})$ and $Q_{hkl}^2$,

$$-\ln(I_{hkl}^{norm}) = -\ln\frac{I_{hkl}(t)}{I_{hkl}^0} = \frac{1}{3} Q_{hkl}^2 \Delta\langle u^2 \rangle(t), \tag{5}$$

where $\Delta\langle u^2 \rangle(t) = \langle u^2 \rangle(t) - \langle u^2 \rangle(t_0)$.

**Supplementary Note 4: Calculations of pair distribution function**

The PDF is calculated on the basis of a reduced distribution function $G(r)$ with adjustments by physical and mathematical constraints. The $G(r)$ can be obtained by the sine transform of the distribution of electron diffraction intensity[4, 5]:

$$G(r) = \frac{2}{\pi} \int_{Q_{min}}^{Q_{max}} Q[S(Q) - 1] \sin(Qr) \, dQ, \tag{6}$$

Where $Q$ is the scattering vector, $r$ represents the separation between atom pairs in real space, and $S(Q)$ is the structure function of the specimen. The relationship between $S(Q)$ and the measured coherent scattering intensity $I(Q)$ with background subtracted is expressed as:

$$S(Q) = 1 + \frac{N \cdot I(Q) - \langle f_e^2(Q) \rangle}{\langle f_e(Q) \rangle^2}, \tag{7}$$

where $N$ is a normalization factor, $\langle f_e^2(Q) \rangle$ and $\langle f_e(Q) \rangle^2$ are the averaged electron

scattering factors for compositions in the specimen[5]. The electron scattering factors of Sb and Se, which are covalently bonded in $Sb_2Se_3$, are provided in literature[6]. The parameter $N$ depending on the scattering factor and intensity is given by

$$N = \frac{\int_{Q_{min}}^{Q_{max}} \langle f_e^2(Q') \rangle dQ'}{\int_{Q_{min}}^{Q_{max}} I_c'(Q') dQ'}. \tag{8}$$

In order to suppress the edge effect introduced by the $Q$ truncation during the integration of Fourier transforms, we implement a damping factor in the calculation of $G(r)$[5]:

$$G(r) = \frac{2}{\pi} \int_{Q_{min}}^{Q_{max}} Q[S(Q) - 1] \sin(Qr) \cdot G(Q) dQ, \tag{9}$$

where $G(Q) = \exp(-0.03 \times Q^2)$ is implemented to suppress the reduced structure function $F(Q) = Q[S(Q) - 1]$, so that $F(Q)$ goes to zero when $Q$ goes to infinity. The PDF $g(r)$ is then obtained by normalizing $G(r)$[4, 5]:

$$g(r) = \frac{G(r)}{4\pi r \rho_0} + 1, \tag{10}$$

where $\rho_0$ is the averaged density of atoms.

**Supplementary Text 5: Contributions of lattice distortion to scattering intensity**

The amplitude of electron scattering in a perfect lattice at zero temperature is given by[7]:

$$\psi(Q) = \sum f_n(Q) e^{-i\vec{Q}\vec{R_n}}, \tag{11}$$

where $\vec{Q}$ is the scattering vector, $\vec{R_n}$ is atomic position of the $n$th atom at equilibrium, and $f_n(Q)$ is the atomic form factor. When the lattice is locally distorted, the atomic position can be written as $\vec{r_n} = \vec{R_n} + \vec{u}$, where $\vec{u}$ represents the small offset of the atomic position; the scattering vector can subsequently be expressed as $\vec{Q} = \vec{G} + \vec{q}$, where $\vec{G}$ is the scattering vector of the nearest Bragg peak, $\vec{q}$ represents the deviation of the scattering vector. Hence, we have

$$\vec{Q} \cdot \vec{r_n} = (\vec{G} + \vec{q}) \cdot (\vec{R_n} + \vec{u}) = \vec{R_n} \cdot \vec{G} + \vec{u} \cdot \vec{G} + \vec{R_n} \cdot \vec{q} + \vec{u} \cdot \vec{q}. \tag{12}$$

$\vec{u}$ and $\vec{q}$ are considered as small quantities, so that their dot product is negligible. Substituting $\vec{Q} \cdot \vec{r_n}$ into the equation (11), we have the following approximation:

$$\begin{aligned}
\psi(Q) &\approx \sum f_n(Q)e^{-i\vec{Q}\vec{R_n}} \cdot e^{-i\vec{G}\vec{u}} \\
&\approx \sum f_n(Q)e^{-i\vec{Q}\vec{R_n}}(1 - i\vec{G}\vec{u}) \\
&= \psi(Q)_{bragg} - i\vec{G}\vec{u}\sum f_n(Q)e^{-i\vec{q}\vec{R_n}}
\end{aligned} \quad (13)$$

Thus, the diffuse scattering signals introduced by the atomic displacement from lattice distortions arising near the Bragg peak are given by

$$\psi(Q)_{diff} = -i\vec{G}\vec{u}\sum f_n(Q)e^{-i\vec{q}\vec{R_n}}. \quad (14)$$

Now we consider the impact of polarons. The atomic displacements of polaron in real space can be phenomenologically described using a Gaussian displacement field model[7] $\vec{u(r)} = A \cdot e^{-r^2/r_p^2(t)} \hat{r}$, where $A$ denotes the displacement amplitude, $r_p(t)$ denotes the radius of polaron, $\hat{r}$ denotes the direction of displacement vector. The diffuse scattering with modulation of the described polaron model can be approximately written as

$$\psi_{diff}(Q) \approx -iA\vec{G}\hat{r}f(Q)\sqrt{\pi}r_p(t)e^{-\frac{q^2 r_p^2(t)}{4}}, \quad (15)$$

where $\sqrt{\pi}r_p(t)e^{-\frac{q^2 r_p^2(t)}{4}}$ is the Fourier transformation of $e^{-r^2/r_p^2(t)}$, $f(Q)$ is the mean atomic form factor. The total scattering intensity produced in a crystal with polarons is then given by

$$\begin{aligned}
I(Q) &= \left|\psi(Q)_{bragg} + \psi(Q)_{diff}\right|^2 \\
&\approx \left|\sum f_n(Q)e^{-i\vec{Q}\vec{R_n}} - iA\vec{G}\hat{r}f(Q)\sqrt{\pi}r_p(t)e^{-\frac{q^2 r_p^2(t)}{4}}\right|^2.
\end{aligned} \quad (16)$$

$\psi(Q)_{bragg}$ contains the crystal shape factor $\sum e^{-iQR_n}$, which approaches $\delta(G)$ in an infinite crystal[8]. Therefore, $I(Q)$ contains no cross term of Bragg scattering and diffuse scattering, we have:

$$I(Q) \approx I_{bragg}(Q) + \pi A^2 f^2(Q)(\vec{G} \cdot \hat{r})^2 r_p^2(t) e^{-\frac{q^2 r_p^2(t)}{2}}. \quad (17)$$

As $\vec{q}$ moves away from the peak center, the intensity of Bragg scattering shows a decay with $1/q^2$ dependence. The normalized differential diffraction intensity obtained by transforming equation (17) is expressed as:

$$\frac{\Delta I}{I_{bragg}} \propto A^2 (\vec{G} \cdot \hat{r})^2 q^2 r_p^2(t) \cdot e^{-\frac{q^2 r_p^2(t)}{2}}. \quad (18)$$

Equation (18) indicates that the differential diffraction intensity modulated by polarons

is dependent on the scattering vector deviation $\vec{q}$ and the polaron radius $r_p(t)$.


\end{document}